\documentclass[prl,aps,twocolumn,showpacs]{revtex4}

\usepackage{graphicx}
\usepackage{dcolumn}
\usepackage{bm}

\begin{document}

\title{Coherence Window in the dynamics of Quantum Nanomagnets}

\author{P. C. E. Stamp$^{1}$, I. S. Tupitsyn$^{2}$}

\affiliation{ $^{1}$ Canadian Institute for Advanced Research,\\
and Physics Dept., University of British Columbia, 6224
Agricultural Rd., Vancouver, B.C., Canada V6T 1Z1\\
$^{2}$  Russian Science Centre "Kurchatov Institute", Moscow
123182, Russia }


\begin{abstract}

Decoherence in many solid-state systems is anomalously high,
frustrating efforts to make solid-state qubits. We show that in
nanomagnetic insulators in large transverse fields, there can be a
fairly narrow field region in which both phonon and nuclear
spin-mediated decoherence are drastically reduced. As examples we
calculate decoherence rates for the $Fe$-8 nanomolecule, for $Ni$
particles, and for $Ho$ ions in $LiHo_xY_{1-z}F_4$. The reduction
in the decoherence, compared to low field rates, can exceed 6
orders of magnitude. The results also give limitations on the
observability of macroscopic coherence effects in magnetic
systems.

\end{abstract}

\maketitle

\vspace{1cm}



Decoherence has emerged as a major challenge, both for fundamental
physics, and for attempts to make solid-state qubits
\cite{preskill}. Experimental decoherence rates are often much
larger than theoretical estimates \cite{mohanty,squbit}- in
complex systems like conductors or superconductors, many
low-energy excitations apart from the usual "oscillator bath"
modes \cite{cal83,weiss} can cause phase decoherence. These
include charged defects, dislocations, paramagnetic and nuclear
spins, external flux and voltage noise, and junction resistance
fluctuations.  The advantages of solid-state systems (stable
circuits, scalability, etc.) cannot be exploited until decoherence
effects are suppressed.

It is often assumed that this is purely an engineering problem.
However this is not true- even completely pure systems have
intrinsic decoherence, and then theory is essential, to see how
this can be quantified and controlled. Here we consider a very
interesting possibility, involving nanomagnetic insulators. These
are often chemically reproducible, so impurities and defects are
kept to a minimum- and electronic decoherence is completely
absent. We show they also have another crucial feature, viz., the
decoherence can be tuned down to very low values. These results
are of wide interest, in the search for magnetic materials showing
spin coherence phenomena.

\vspace{2mm}


{\bf 1. Intrinsic Decoherence in a nanomagnetic "qubit"}: At low
$T$ the spin Hamiltonian of many large-spin nanomagnetic systems
(magnetic molecules, rare earth ions, or nanomagnetic particles)
reduces to a 2-state form ${\cal H}_o(\hat{\tau}) = (\Delta_o
\hat{\tau}_x + \epsilon_o \hat{\tau}_z)$, with the Pauli spin
$\hat{\tau}$ acting on the 2 lowest spin levels
\cite{wwRev,tupBB}. The spin gap $E_G$ to the next levels is
typically $\sim 5-10~K$, and the 2-state picture is valid at
energies $\ll E_G$. At higher energies the individual electronic
spins $\{ {\bf s}_j \}$ are assumed to be locked into a nanospin
${\bf S}$ (so that ${\bf S} = \sum_j {\bf s}_j$) by strong
exchange or superexchange interactions. We assume henceforth an
"easy $\hat{z}$-axis" nanomagnet; then the 'bias' energy
$\epsilon_o = g \mu_B S_zH_o^z$. When $\epsilon_o = 0$, the
splitting $\Delta_o$ between the 2 "qubit" states $\vert -
\rangle, \vert + \rangle$ (bonding and anti-bonding eigenstates of
${\cal H}_o(\hat{\tau})$) is produced by tunneling between 2
potential wells, each well having a "small oscillation" energy
$\Omega_o$; typically $\Omega_o \sim E_G$. We define the states
$\vert \uparrow \rangle,\vert \downarrow \rangle$ (eigenstates of
$\hat{\tau}_z$) by $\vert \pm \rangle = [\vert \uparrow \rangle
\pm \vert \downarrow \rangle]/\sqrt{2}$. If the total nanomagnetic
spin ${\bf S}$ is not too small, these states correspond roughly
to semiclassical spin coherent states \cite{auerbach}, having
orientations ${\bf n}_{\sigma}$ (here $\sigma =
{\uparrow},{\downarrow}$), which depend on both the internal
anisotropy field of the nanomagnet, and any transverse external
field ${\bf H}_o^{\perp}$. The splitting $\Delta_o$ depends
sensitively on ${\bf H}_o^{\perp}$.

The intrinsic decoherence in insulating nanomagnets comes from
entanglement of the nanomagnetic spin wave function with that of
the nuclear spins and phonons \cite{PS96}. Both of these couplings
are well understood. The nuclear spins $\{ {\bf I}_k \}$ couple to
the electronic spins $\{ {\bf s}_j \}$ in ${\bf S}$ via individual
hyperfine couplings $A_{\alpha \beta}^{jk} s_j^{\alpha}
I_k^{\beta}$, and the phonons couple to ${\bf S}$ via
magnetoacoustic (spin-phonon) interactions \cite{vill96}. Without
yet specifying the precise form of the hyperfine couplings (which
of course depend on the system being studied) we can define quite
generally a vector $\vec{\omega}_k^{\parallel}$ which specifies
the net effect on the $k$-th nuclear spin ${\bf I}_k$, of all the
hyperfine fields coming from the individual electronic spins in
${\bf S}$:
\begin{equation}
\vec{\omega}_k^{\parallel} \;\equiv\; \omega_k^{\parallel}
\hat{\it l}_k^{\alpha} \;=\; {1 \over 2}  \sum_j A_{\alpha
\beta}^{jk} (\langle s_j^{\alpha} \rangle^{\uparrow} - \langle
s_j^{\alpha} \rangle^{\downarrow})
 \label{omega}
\end{equation}
Here $\hat{\it l}_k$ is a unit vector in the direction of this
total field, with components $\hat{\it l}_k^{\alpha}$, and
$\langle s_j^{\alpha} \rangle^{\sigma}$ is the expectation value
of ${\bf s}_j$ when ${\bf S} \rightarrow S{\bf n}^{\sigma}$. We
see that when ${\bf S}$ flips from $S{\bf n}_{\uparrow}$ to $S{\bf
n}_{\downarrow}$, the energy change is just
$2\omega_k^{\parallel}\hat{\it l}_k \cdot {\bf I}_k$, ie., we have
a diagonal coupling $\hat{\tau}_z \omega_k^{\parallel} \hat{\it
l}_k \cdot \hat{\bf I}_k$ between the qubit and ${\bf I}_k$. One
can also write this coupling as $\hat{\tau}_z \xi_z$, where $\xi_z
= \omega_k^{\parallel} \hat{\it l}_k \cdot \hat{\bf I}_k$ acts as
an extra bias field, in addition to the external bias field
$\epsilon_o$. Note that the weak interaction between nuclear spins
(typically dipolar) causes slow spin diffusion between them. The
net effect of this is to make $\xi_z \rightarrow \xi_z(t)$, ie.,
the nuclear bias field acting on the qubit fluctuates in time even
when the qubit is frozen. We shall argue below that we can neglect
this fluctuation in strong transverse fields.

The ${\bf I}_k$ also couple to the external field ${\bf H}_o$,
with Zeeman coupling $\omega_k^{\perp} \hat{\it m}_k \cdot
\hat{\bf I}_k$, where
\begin{equation}
\omega_k^{\perp} \hat{\it m}_k = g_k^N \mu_N {\bf H}_o
 \label{zeeman}
\end{equation}
and $\hat{\it m}_k$ is a unit vector along ${\bf H}_o$.

If we now take these terms together, we can write the interaction
Hamiltonian between the nanospin ${\bf S}$ and the nuclear spins
$\{ {\bf I}_k \}$ in the form ${\cal H}_{NS} = {\cal
H}_o(\hat{\tau}) + V(\hat{\tau}, {\bf I}_k)$, where
\begin{equation}
V =  \hat{\tau}_z  \sum_k \omega_k^{\parallel} \hat{\it l}_k \cdot
\hat{\bf I}_k  \;+\; \sum_k \omega_k^{\perp} \hat{\it m}_k \cdot
\hat{\bf I}_k
 \label{Ham}
\end{equation}
This form is particularly useful for quantifying the decoherence
from the nuclear spins. When we come to particular examples we
will specify the couplings in (\ref{Ham}).

The magnetoacoustic interaction between the qubit coordinate
$\hat{\tau}$ and the phonon coordinate $x_q$ is dominated in
nanomagnets \cite{vill96,PS96} by a non-diagonal term
$\hat{\tau}_x\sum_q c_q^{\perp} x_q$. This term has strength
\begin{equation}
c_q^{\perp} x_q \sim S \Omega_o (\omega_q / \theta_D)^{1/2},
\label{s-ph}
\end{equation}
where $\theta_D$ is the Debye energy, $\omega_q = q c_s$ and $c_s$
is the sound velocity.

\vspace{2mm}

Our basic idea is as follows. The allowed nuclear spin bath
states, when the qubit is in some given state, have a density of
states which typically has Gaussian lineshape \cite{lineshape},
with a halfwidth $E_o$; in terms of the $\{ \omega_k^{\parallel}
\}$ defined above, $E_o$ is given trivially by
\begin{equation}
E_o^2 = \sum_k {I_k + 1 \over 3I_k} (\omega_k^{\parallel}I_k)^2
 \label{E_o}
\end{equation}
On the other hand the acoustic phonon energy scale is the Debye
energy $\theta_D$. Now in a nanomagnetic system $E_o/\theta_D$ can
be $ \lesssim 10^{-4}$, suggesting the following tactic for
suppressing decoherence. If we tune $\Delta_o$ so that $\theta_D
\gg \Delta_o \gg E_o$, then we will be in a "coherence window", in
which decoherence will be at a minimum because the qubit dynamics
is too slow to disturb most phonons, but too fast for the nuclear
spins to react.

\vspace{2mm}

{\it Decoherence Rates:} To substantiate this idea, we generalise
the low field ($\Delta_o < E_o$) calculations of nanomagnetic
dynamics \cite{PS96}, where incoherent tunneling relaxation is
found, to the high-field regime $\Delta_o \gg E_o$. Because at
high field the couplings $\{ \omega_k^{\parallel} \}$ of the
nuclear spins to the qubit are $\ll$ the nuclear Zeeman couplings
$\{ \omega_k^{\perp} \}$, and also $\{ \omega_k^{\parallel} \} \ll
\Delta_o$, this dynamics can be solved perturbatively
\cite{PS00,castro93}. We expand about the bare qubit Hamiltonian
(\ref{Ham}) to 2nd order in $\omega_k^{\parallel}/\Delta$
(assuming $\epsilon_o = 0$ for simplicity) to get
\begin{eqnarray}
H_{NS} &=& [ \Delta_o \hat{\tau}_x + \sum_k \omega_k^{\perp}
\hat{\it m}_k \cdot {\bf I}_k]   \nonumber \\
&+& \hat{\tau}_x \; \sum_{kk'} {\omega_k^{\parallel}
\omega_{k'}^{\parallel} \over 2 \Delta_o} (\hat{\it l}_k \cdot
{\bf I}_k) (\hat{\it l}_{k'} \cdot {\bf I}_{k'})  \nonumber \\
&+& \; O((\omega_k^{\parallel})^4/\Delta_o^3)
 \label{12a}
\end{eqnarray}

The decoherence time $\tau_{\phi}$ is defined as the
characteristic time for decay of the off-diagonal density matrix
element, starting in state $\vert \uparrow \rangle$. In the
present case we calculate this matrix element as a path integral
over pairs of qubit trajectories $\tau_z(t), \tau_z(t')$ (each
taking values $\pm 1$, with occasional flips between these
values), weighted by an influence functional $F \left[ \tau_z(t),
\tau_z(t') \right]$ which incorporates the interactions
\cite{weiss}. The contribution from the 2nd term in (\ref{12a}) to
this functional is \cite{castro93,PS00}:
\begin{equation}
\ln F= -\sum_k { (\omega_k^{\parallel} )^2 \over 8\hbar^2 } \vert
\int_0^t  ds  e^{{i \over \hbar} \omega_k^{\perp} s } \big[
\tau_z(s)- \tau_z'(s) \big] \vert ^2
 \label{lnF}
\end{equation}
From this result we can then use standard techniques developed for
the spin-boson model \cite{weiss}, to find a contribution
$\gamma_{\phi}^{\kappa}$ to the dimensionless decoherence rate
$\gamma_{\phi} = 1/\tau_{\phi} \Delta_o$ (the commonly used
'decoherence quality factor' \cite{squbit} is just $Q_{\phi} =
\pi/\gamma_{\phi}$). We find
\begin{eqnarray}
\gamma_{\phi}^{\kappa} &=& \sum_{kk'} \sqrt{{(I_k + 1)(I_{k'} + 1)
\over 9I_k I_{k'}}} {\omega_k^{\parallel} \omega_{k'}^{\parallel}
I_k I_{k'} \over 2\Delta_o^2}  \nonumber \\
&=& {1 \over 2} \left ( {E_o \over \Delta_o}
\right )^2
\label{kapp2}
\end{eqnarray}
which decreases rapidly with increasing qubit operating frequency
$\Delta_o$.

There are 2 other contributions to $\gamma_{\phi}$ coming from the
nuclear spins \cite{PS96,PS00}. First, in writing (\ref{Ham}) we
omitted a renormalisation of the tunneling matrix element caused
by the coupling to the nuclear spins \cite{topoD}, which in fact
describes the nuclear spin transitions induced directly by
electronic spin flips. This adds a contribution
$\gamma_{\phi}^{\lambda}$ to $\gamma_{\phi}$, given by
$\gamma_{\phi}^{\lambda} = {1 \over 2} \sum_k \vert \vec{\alpha}_k
\vert^2$, where $\vert \vec{\alpha}_k \vert = \pi \vert
\omega_k^{\parallel} \vert /2 \Omega_o$. However when $\Delta_o
\gg E_o$, the ratio $\gamma_{\phi}^{\lambda}/
\gamma_{\phi}^{\kappa} \sim O(\Delta_o^2/\Omega_o^2) \ll 1$ (the
usual WKB reduction of the tunneling amplitude), ie., this term
can always be neglected to first approximation. Second, we have
neglected the instrinsic nuclear spin diffusion dynamics, caused
by internuclear interactions. In low fields, when the tunneling is
slow, and $\Delta_o \ll E_o$, this intrinsic nuclear dynamics
renders the tunneling dynamics incoherent
\cite{PS96,wwNS,thomas,giraud}. However when $\Delta_o \gg E_o$,
the nuclear fluctuations are very slow compared to the qubit
dynamics (of frequency $\Delta_o$); they then add a 'noise'
contribution $\gamma_{\phi}^N \sim N/\pi \Delta_o T_2$ to
$\gamma_{\phi}$, where $N$ is the number of nuclear spins in each
molecule \cite{PS00}. Typically $T_2^{-1} \sim 10-100~Hz$ at low
$T$ in magnetic molecules \cite{morello02}), whereas we are
interested in $\Delta_o \sim GHz$ (see below); thus
$\gamma_{\phi}^N$ will be very small.

Finally we include the phonon contribution to $\gamma_{\phi}$. The
solution to the spin-boson problem for non-diagonal coupling to
phonons \cite{weiss} gives a contribution $\gamma_{\phi}^{ph}$ of
form \cite{PS96}:
\begin{equation}
\gamma_{\phi}^{ph} = [(S \Omega_o \Delta_o)^2/\Theta_D^4]
\coth(\Delta_o / k_B T)
 \label{Dph}
\end{equation}
which increases rapidly with $\Delta_o$.

The phonon and nuclear spin decoherence mechanisms act
independently- thus we can get a simple estimate for the optimal
decoherence rate $\gamma_{\phi}^{min}$ by summing the two dominant
contributions (\ref{kapp2}) and (\ref{Dph}), and minimizing their
sum $\gamma_{\phi}^{\kappa} + \gamma_{\phi}^{ph}$ with respect to
$\Delta_o$, assuming $k_BT < \Delta_o$. This gives
\begin{equation}
\gamma_{\phi}^{min} \approx \sqrt{2} S \Omega_o E_o / \theta_D^{2}
\label{g-min}
\end{equation}
at an optimal tunneling splitting $\Delta_o^{opt}$:
\begin{equation}
\Delta_o^{(opt)} \approx \theta_D (E_o / \sqrt{2} S \Omega_o)^{1/2}.
\label{D-opt}
\end{equation}
We see that decoherence is minimized for a given $S$ by making
$E_o$ and $\Omega_o$ small, and $\theta_D$ large, within the
constraint that $\Omega_o \gg \Delta_o > k_BT$. If $k_BT >
\Delta_o$ we get a different (less favorable) answer.

These simple results actually give reasonably accurate results
when compared with numerical calculations on real systems, as we
now see.

\vspace{4mm}


{\bf 2. Three Examples}: We present quantitative results for the
decoherence rates in 3 different materials. We give most details
for the $Fe$-8 example, to illustrate our method.

\vspace{2mm}

{\bf (i) The $Fe$-8 molecule}: This well-characterized molecule
\cite{wwRev} behaves below $\sim 10~K$ as an electronic spin-$10$
system, with biaxial effective Hamiltonian \cite{wwRev,tupBB}
${\cal H}_o({\bf S}) \sim -D\hat{S}_z^2 + E\hat{S}_x^2 +
K_4^{\perp}(\hat{S}_+^4 + \hat{S}_-^4) - g\mu_B {\bf H}_{\perp}
\cdot {\bf S}$, with $D/k_B = 0.23~K$, $E/k_B = 0.094~K$, and
$K_4/k_B = -3.28 \times 10^{-5}~K$. The small oscillation
frequency $\Omega_o = 2SC_4\sqrt{DE} \sim 4.6~K$, where $C_4 =
1.56$ includes the effects of the $K_4$ term; $T$-independent
tunneling dynamics appears below $\sim 0.4~K$. The tunneling
amplitude $\Delta_o({\bf H}_o^{\perp})$ is trivially determined by
diagonalisation of ${\cal H}_o({\bf S})$ (Fig. 1), as are the spin
orientations ${\bf n}^{\uparrow},{\bf n}^{\downarrow}$ defined
previously. At a critical transverse field $H_c$ the barrier is
destroyed and ${\bf n}^{\uparrow},{\bf n}^{\downarrow}$ merge.
When ${\bf H}_o^{\perp}$ is along the hard $\hat{x}$-axis (so the
azimuthal angle $\phi = 0$ or $\pi$), one has $g \mu_B H_c = 2 S
(D + E)$ for this Hamiltonian, giving $H_c \sim 4.8~T$. We find by
comparing exact diagonalisation with semiclassical calculations,
that the latter are accurate up to ${\bf H}_o^{\perp} \sim
3.3-4.3~T$ (for $\phi = 90^o,0^o$ respectively).

The hyperfine interactions between the 8 $Fe^{+3}$ (spin $5/2$)
ions and the 205 nuclear spins in the molecule (213 if $^{57}Fe$
isotopes are substituted for $^{56}Fe$ nuclei) are of 2 kinds. The
$Fe$ electronic spins interact with any $^{57}Fe$ ions via contact
hyperfine interactions, which we assume to be the same as for
$Fe^{+3}$ ions in similar materials \cite{rado}. On the other hand
the hyperfine couplings $A_{\alpha \beta}^{jk} s_j^{\alpha}
I_k^{\beta}$ to all the other nuclear spins are thought to be
dominated by purely dipolar terms \cite{wwNS}.

\begin{figure}[h]
\centering
\vspace{-2.0cm}
\includegraphics[scale=0.35]{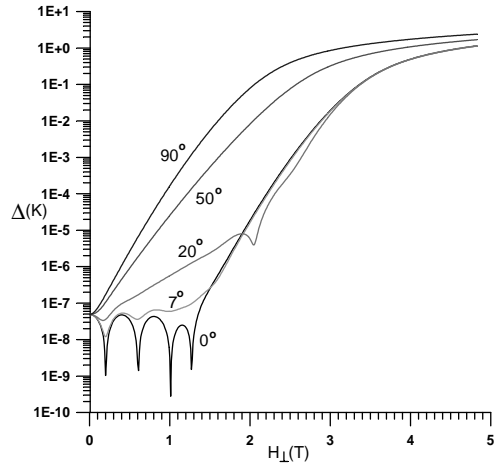}
\vspace{-1.5cm}
\caption{The tunneling splitting $\Delta_o({\bf
H}_{\perp})$ for the $Fe$-8 molecule, for transverse field ${\bf
H}_{\perp}$ in the $\hat{x}\hat{y}$ plane, at azimuthal angle
$\phi$ from the in-plane hard axis. We use parameters given in the
text \cite{wwRev,tupBB} for the molecular spin Hamiltonian ${\cal
H}_o({\bf S})$.} \label{fig:fig1}
\end{figure}

Using the known nuclear positions and moments, the hyperfine
interactions and the nuclear Zeeman couplings are then quantified
numerically, to find the $\{ \omega_k^{\parallel} \}$ and hence
$E_o$ (see Fig. 2), and also the $\{ \omega_k^{\perp} \}$. In
spite of the large number of protons, $E_o$ is quite small
(particularly when we substitute $^2H$ nuclei for $^1H$), making
$Fe$-8 a reasonable candidate for coherent dynamics in high
fields. One may also calculate the nuclear dynamics \cite{fine},
but here we simply note that NMR experiments \cite{morello02}
indicate that at low $T$, $T_2 \sim 10-30~msec$ for  $^1H$ nuclei
in magnetic molecules. We then calculate numerically the different
nuclear spin contributions  to $\gamma_{\phi}$ (ie., the dominant
contribution (\ref{kapp2}), plus the smaller contributions
$\gamma_{\phi}^{\lambda}$ and $\gamma_{\phi}^{N}$), and also the
phonon contribution $\gamma_{\phi}^{ph}$, as a function of ${\bf
H}_o^{\perp}$.

The results are shown for high fields in Fig. 3. Actually almost
all experiments on the quantum dynamics of $Fe$-8 have been done
in the regime $\Delta_o < E_o$, ie., where we expect nuclear spins
to cause incoherent tunneling \cite{PS96}- which is what is found
experimentally \cite{wwNS,morello02,sangr97,thomas}. However once
$\Delta_o$ exceeds $E_o$, all nuclear spin decoherence should fall
off very fast. The nuclear noise contribution $\gamma_{\phi}^N <
10^{-7}$ once $\vert {\bf H}_o^{\perp} \vert
> 3.2~T$ (for $\phi = 0^o$) or $> 2~T$ (for $\phi > 90^o$), so we
can safely ignore it. The estimate (\ref{g-min}) then gives
$\gamma_{\phi}^{min} \approx 5.8 \times 10^{-5}$ at a
$\Delta^{opt}_o \approx 0.14~K$, when $\phi=0$ and $H_o^{\perp} =
3.45~T$ (assuming the optimal set of nuclear isotopes, with $^2H$
instead of $^1H$, no $^{57}Fe$, etc). In Fig. 3 we calculate
$\gamma_{\phi}$ numerically, adding all nuclear spin and
phonon contributions. For $\phi =0$ we get a numerical value
$\gamma_{\phi}^{min} = 6 \times 10^{-5}$, at a $\Delta_o^{opt} =
0.135~K$ when $H_o^{\perp} = 3.45~T$; thus the estimate works well
for $\phi = 0$. The further reduction of $\gamma_{\phi}^{min}$ to
$ \sim 2.4 \times 10^{-6}$, when $\phi \rightarrow 90^o$
($H_o^{\perp} = 2.93~T$), comes from decreases in both $\Omega_o$
and the effective $S$ (since both the tunneling barrier and the
change $\vert {\bf S}^{\uparrow} - {\bf S}^{\downarrow} \vert$
decrease). Fig. 3 clearly illustrates the window for coherent
dynamics which opens up at high transverse fields ($H_o^{\perp}
\sim 2.9-3.4~T$, depending on $\phi$) in $Fe$-8.

\begin{figure}[h]
\centering
\vspace{-1.8cm}
\includegraphics[scale=0.35]{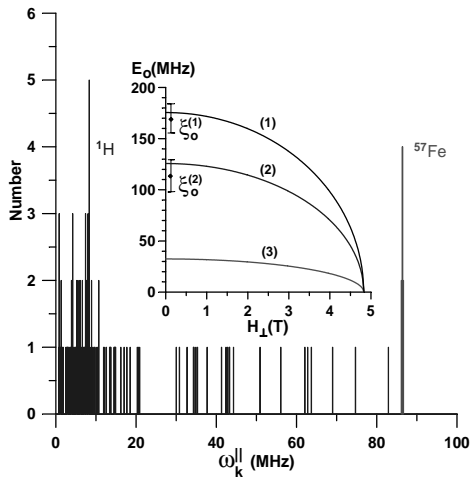}
\vspace{-1.8cm} \caption{A histogram of the calculated couplings
$\{\omega_k^{\parallel} \}$ between the $Fe$-8 spin-10 and both
$^{1}H$ nuclei and $^{57}Fe$ nuclei, binned in $0.1~MHz$ intervals
(the couplings to the other nuclear species are not displayed).
Apart from the core polarisation coupling to $^{57}Fe$ nuclei, all
hyperfine couplings are assumed dipolar, and calculated using the
known positions of each nucleus \cite{camdata}. We assumed
Hartree-Fock wave-functions for each $Fe^{+3}$ ion
\cite{rado}(correcting the results of ref. \cite{rose}, which used
an incorrect coordinate system for the molecule). The insert shows
the variation with ${\bf H}_o^{\perp}$ of the nuclear multiplet
linewidth parameter $E_o$ (see text). Curve (2) has naturally
occurring isotopic concentrations. Curve (1) substitutes $^{57}Fe$
for $^{56}Fe$, and curve (3) substitutes $^{2}H$ for $^{1}H$.
Previous measurements of the "hole-width" parameter $\xi_o$, which
has $E_o$ as an upper bound (taken from ref. \cite{wwNS}) are also
shown for 2 of these cases.} \label{fig:fig2}
\end{figure}


{\bf (ii) $Ni$-based particles}: Consider now a pure $Ni$ particle
at low $T$, with $\Omega_o = 1~K$ and spin $S$, coupled to a
substrate with $\theta_D = 300~K$. The concentration of $^{59}Ni$
nuclei (spin-$1/2$) is $x_{59} \sim 0.01$ in natural $Ni$, and the
hyperfine coupling $\omega_k^{\parallel}\rightarrow \omega_o \sim
1.4~mK$. From (\ref{g-min}) we have $\gamma_{\phi}^{min} \sim
SE_o/10^5 \sim O (S^{3/2} \omega_o x_{59}^{1/2}/10^5)$, since $E_o
\sim \omega_o N^{1/2}$, where $N \sim S$ is the number of nuclear
spins. Then in natural $Ni$, $\gamma_{\phi}^{min}$ reaches unity
once $S \sim 2.6 \times 10^4$, for a $\Delta_o^{(opt)} \sim
0.2~K$. Isotopic purification by a factor $10^2$ (a major
undertaking!) would reduce $\gamma_{\phi}^{min}$ by only a factor
of 10. Note that $Ni$ has an unusually low concentration of
nuclear spins, with rather weak hyperfine couplings. Thus this
example teaches us that to see any macroscopic coherence effects
in tunneling magnets for $S > 10^5$ will require almost complete
isotopic purification (there are no known magnetic species not
having at least one natural isotope with non-zero nuclear spin).
Note further that we only consider here intrinsic decoherence from
phonons and nuclear spins inside the particle- we have ignored
electronic decoherence, which certainly exists in $Ni$.

\begin{figure}[h]
\centering
\vspace{-1.8cm}
\includegraphics[scale=0.35]{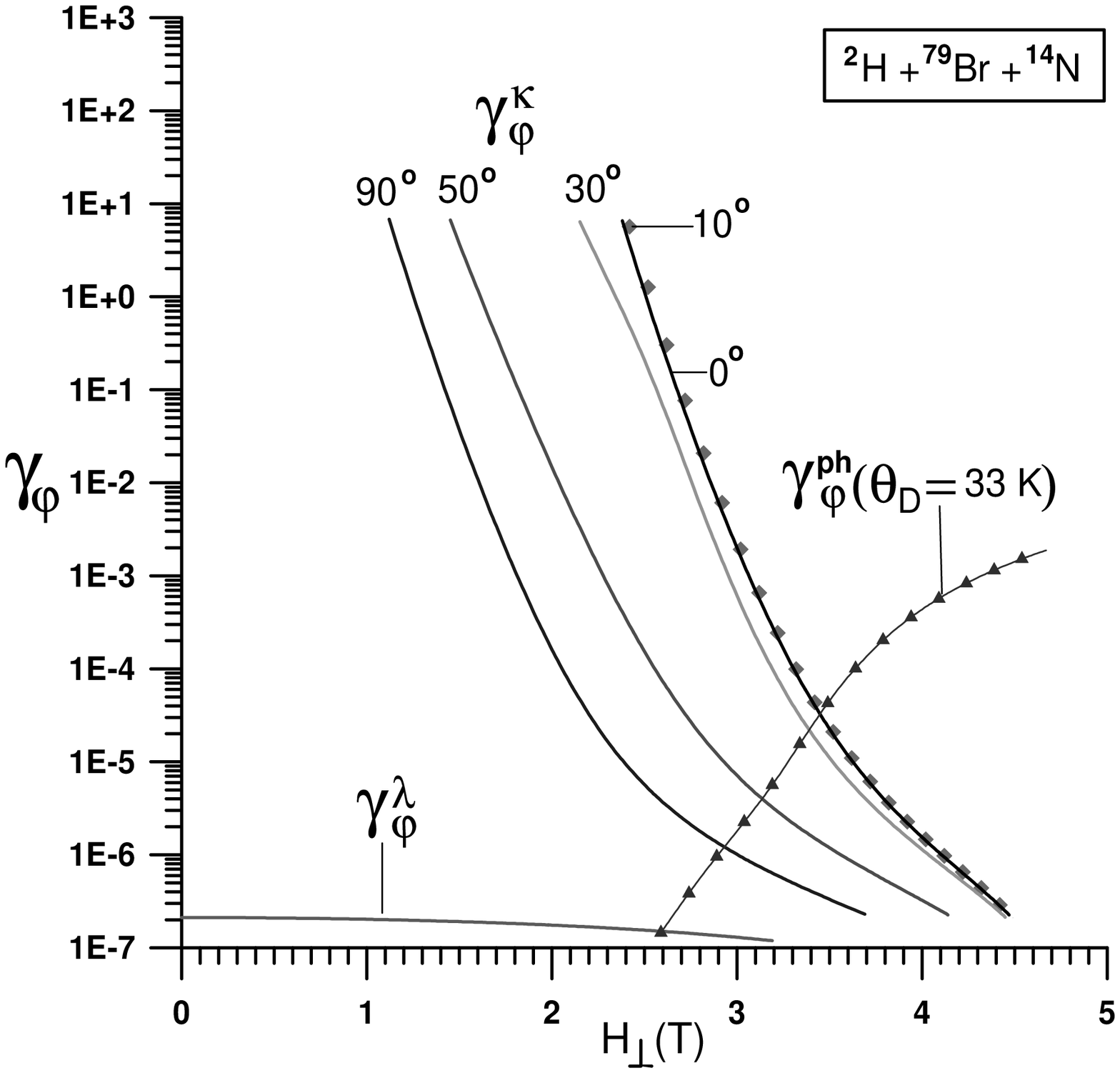}
\vspace{-1.8cm} \caption{Calculated behaviour of
$\gamma_{\phi}({\bf H}_{\perp})$ for an $Fe$-8 molecule, using the
results for $\Delta_o({\bf H}_o^{\perp}$, the $\{
\omega_k^{\parallel}({\bf H}_{\perp}) \}$, and $E_o$ (Fig. 1), and
also evaluating $\Omega_o ({\bf H}_{\perp})$ as a function of
${\bf H}_{\perp}$. We show the individual contributions from
$\gamma_{\phi}^{\kappa}$ for different field orientations in large
transverse fields, for $\gamma_{\phi}^{\lambda}$ (shown for $\phi
= 0^0)$, and for the phonon contribution $\gamma_{\phi}^{ph}$,
assuming a Debye energy $\theta_D = 33~K$ (the small contribution
$\gamma_{\phi}^N$ is not shown). Results are shown for the {\it
optimal} distribution of nuclear isotopes, ie., with $^2H$
substituted for $^1H$, and only $^{56}Fe$, $^{79}Br$, $^{14}N$ and
$^{16}O$ species in the molecule.}
 \label{fig:fig3}
\end{figure}


{\bf (iii) $Ho$ ions}: Not all systems will have an optimal
intrinsic decoherence described by (\ref{g-min}). Consider the
$LiHo_xY_{1-x}F_4$ system \cite{aeppli}, in which sharp absorption
lines are seen \cite{giraud,ghosh02} at low $x$. The $Ho^{+3}$
ions (with spin $=8$) have a lowest doublet state, with splitting
$\Delta_o \propto \vert {\bf H}_o^{\perp}\vert^2$ in a transverse
field- only one other state, at an energy $\Omega_o \sim 10.6~K$,
is important for the low-energy physics (all other electronic
spins levels are at energies $\gtrsim 100~K$). The new feature
here, which renders (\ref{g-min}) inapplicable, is a very strong
hyperfine coupling $\omega_{Ho}^{\parallel} = 0.039~K$ to the
$I=7/2$ $Ho$ nuclear spin- whereas the couplings to the
"satellite" $Li,Y$, and $F$ nuclear spins are very weak. Thus
instead of having a Gaussian lineshape, the hyperfine multiplet
has a "toothcomb" structure with 15 $Ho$ lines spaced in intervals
of $0.039~K$, each weakly broadened (by less than $1~mK$) by the
other nuclei. The quantity $E_o$ is then not well-defined
\cite{lineshape}. However one may instead just calculate the
coupled $Ho$-nuclear spin dynamics in 2nd-order perturbation
theory, since $\Omega_o \gg$ all hyperfine couplings, and all $Ho$
transitions go via the single intermediate level. We then find the
optimal strategy is to (a) freeze the $Ho$ nuclei by cooling to
$k_BT \ll 0.039~K$, and (b) make ${\bf H}_o^{\perp}$ large enough
so that precessional decoherence from the satellite nuclei is
eliminated, but small enough so that $Ho$ nuclei are weakly
excited. We find that at $T = 3~mK$, tuning of $\Delta_o$ to
$30~mK$ (using ${\bf H}_o^{\perp} \sim 0.3~T$) gives a
contribution $\gamma_{\phi}^{min} \sim 1.8 \times 10^{-5}$ from
the $Ho$ nuclei (phonons giving a contribution $< 10^{-9}$ here).
Some residual decoherence also comes from the 'satellite' nuclear
spins- in this context it is interesting that $Ho^{+3}$ ions can
be prepared in hosts with almost no other nuclear spins (eg.,
$CaWO_4$, with isotopically purified $W$; see \cite{giraud}). Thus
we can expect a reduction of $\gamma_{\phi}^{min}$ to very low
levels, limited only by a weak phonon effect, external noise, and
any spin impurities.

\vspace{4mm}


{\bf 3. Discussion}: In insulating systems where nuclear spin
decoherence is suppressed by a transverse field, decoherence
optimisation is well described by (\ref{g-min}). This means having
a small $S$, small $E_o$, and a "stiff" system (high $\theta_D$).
Surprisingly it also means low $\Omega_o$, ie., weak magnetic
anisotropy. By taking all these measures, very low values of
$\gamma_{\phi}$ can be attained.

Our results are thus good news for coherence in small nanomagnets,
and magnetic qubits. In most materials a correct choice of
parameters requires strong transverse fields, but in systems like
$LiHo_xY_{1-x}F_4$ with one very strong hyperfine coupling
$\omega_o^{\parallel}$, and otherwise weak or non-existent ones,
it pays to keep $\Delta_o$ low (ie., not too strong fields), and
to have $kT \ll \omega_o^{\parallel}$ (to suppress thermal nuclear
spin noise).

For genuinely macroscopic superpositions of magnetic states, the
result (\ref{g-min}) is not such good news. Although one may stave
off decoherence for large $S$ by isotopic purification of nuclear
spins, our $Ni$ example shows this tactic reaches its limit once
$S \sim 10^5-10^6$.

So far experiments on nanomagnets have concentrated on field
ranges where $\Delta_o < E_o$, and so incoherent tunneling is
observed- any experiments on, eg., $Fe$-8 in the range $\vert {\bf
H}_o^{\perp} \vert \sim 3-4~T$ would be of great interest. In
almost all experiments so far, on molecules or rare earth ions,
collective tunneling caused by inter-spin dipolar interactions
complicate the interpretation \cite{aeppli,giraud,ghosh02}. A
clean observation of coherence, with a measurement of
$\gamma_{\phi}$, will thus involve manipulation at microwave
frequencies, on a properly isolated single system.

We thank G.A. Sawatzky for useful discussions, and NSERC, the
CIAR, and grant number NS-1767.2003.2 in Russia, for support.


\end{document}